\documentclass{article}
\usepackage{graphicx}
\usepackage{amsmath}
\usepackage{amssymb}
\usepackage{authblk}
\usepackage{natbib}
\title{The Spatial Whitham Equation}
\author[1]{John D.~Carter\thanks{Corresponding author: carterj1@seattleu.edu}}
\author[2]{Diane Henderson}
\author[3]{Panayotis Panayotaros}

\affil[1]{Mathematics Department, Seattle University, 901 12th Ave, Seattle, WA 98122, USA}
\affil[2]{Department of Mathematics, Penn State University, 218 McAllister Building, University Park, PA 16802}
\affil[3]{Instituto de Investigaciones en Matem\'aticas Aplicadas y en Sistemas, Universidad Nacional Aut\'onoma de M\'exico, Apdo. Postal 20-716, 01000 M\'exico D.F., M\'exico}

\begin{document}
\maketitle

\begin{abstract}

The Whitham equation is a nonlocal, nonlinear partial differential equation that models the temporal evolution of spatial profiles of surface displacement of water waves.  However, many laboratory and field measurements record time series at fixed spatial locations.  In order to directly model data of this type, it is desirable to have equations that model the spatial evolution of time series.  The spatial Whitham equation, proposed as the spatial generalization of the Whitham equation, fills this need.  In this paper, we study this equation and apply it to water-wave experiments on shallow and deep water.

We compute periodic traveling-wave solutions to the spatial Whitham equation and examine their properties, including their stability.  Results for small-amplitude solutions align with known results for the Whitham equation.  This suggests that the systems are consistent in the weakly nonlinear regime.  At larger amplitudes, there are some discrepancies.  Notably, the spatial Whitham equation does not appear to admit cusped solutions of maximal wave height.  In the second part, we compare predictions from the temporal and spatial Korteweg-deVries and Whitham equations with measurements from laboratory experiments.  We show that the spatial Whitham equation accurately models measurements of tsunami-like waves of depression and solitary waves on shallow water. Its predictions also compare favorably with experimental measurements of waves of depression and elevation on deep water.  Accuracy is increased by adding a phenomenological damping term.  Finally, we show that neither the spatial nor the temporal Whitham equation accurately models the evolution of wave packets on deep water.

\end{abstract}

\section{Introduction}
\label{Introduction}

We investigate the spatial Whitham (sWhitham) equation, proposed by \citet{Trillo}, for gravity waves propagating on the surface of water.  This equation was motivated by the success of the temporal Korteweg-de Vries (tKdV) and temporal Whitham (tWhitham) equations in modeling the evolution of unidirectional waves in shallow water.  The tWhitham equation uses a more accurate linear dispersive term than the tKdV equation and therefore may be useful for depths ranging from shallow to deep.  The tWhitham equation can also approximate some singularity formation effects, see \citet{Whitham} and \citet{WhithamCusp}.  Comparisons with experimental data suggest the tWhitham equation improves on the tKdV equation in the small-amplitude long-wave regime, see \citet{Trillo} and \citet{WhithamComp}.

A limitation of temporal models is that they describe how a given surface displacement profile evolves in time.  However, many field and laboratory experiments provide temporal profiles of the surface displacement at fixed spatial locations and ask how these temporal profiles evolve in space.  This problem arises in many wave phenomena, e.g.~in optics, and there is a long use of ``spatial evolution'' equations that describe how a signal, e.g.~an optical image, changes along some direction in space.  Despite their motivation, spatial equations have not been used systematically for nonlinear dispersive water wave models.  Herein, we examine a recently proposed model, the sWhitham equation of \citet{Trillo}, and show evidence that this equation can improve on predictions of the tWhitham, tKdV, and sKdV equations.  We consider this equation mathematically, numerically, and experimentally and compare its properties and predictions with those of the tKdV, tWhitham, and sKdV equations.  Our results yield new information on the properties of the sWhitham equation and its strengths and limitations as a model of unidirectional water waves.

The motion of a two-dimensional, irrotational, incompressible, inviscid fluid on a horizontal, impermeable bed can be modeled by

\begin{subequations}
\begin{equation}
 \phi_{xx}+\phi_{zz} = 0,~~~  \text{ for }  - h_0 < z < \eta,
\end{equation}
\begin{equation}
 \phi_{t} +\frac{1}{2}|{\nabla\phi}|^{2}+g\eta = 0 ,\: ~~~ \text{at}  \: z = \eta,
\end{equation}
\begin{equation}
 \eta_{t}+\eta_{x}\phi_{x} = \phi_{z},~~~\text{ at } z = \eta,
\end{equation}
\begin{equation}
\phi_z=0,~~~ \text{ at } z=-h_0,
\end{equation}
\label{WWProb}
\end{subequations}
where $\phi(x, z, t)$ represents the velocity potential of the fluid, $\eta(x, t)$ represents the surface displacement of the fluid, $g$ represents the acceleration due to gravity, $h_0$ represents the uniform depth of the fluid at rest, and $x$, $z$, $t$ represent the horizontal, vertical, and temporal coordinates, respectively.  See \citet{johnson} for details of this system.  Since this system is a nonlinear free-boundary problem, approximate models are often used.  Linear theory gives the dispersion relation

\begin{equation}
    \omega^2=g\kappa\tanh(\kappa h_0),
    \label{WWPDR}
\end{equation}
where $\kappa$ and $\omega$ are the wavenumber and angular frequency of the linear wave, respectively.

The dimensional temporal \citet{kdv} (tKdV) equation,

\begin{equation}
    \eta_t+\sqrt{gh_0}~\eta_x+\frac{1}{6}h^2\sqrt{gh_0}~\eta_{xxx}+\frac{3}{2h_0}\sqrt{gh_0}~\eta\eta_x=0,
    \label{KdVDimensional}
\end{equation}
is a small-amplitude, long-wave approximation of (\ref{WWProb}).  The tKdV equation has been well studied mathematically (e.g.~\citet{miles,AS,LannesBook}) and experimentally (e.g.~\citet{russell,zabusky,h,HS}).  It has dispersion relation

\begin{equation}
    \omega_{\text{KdV}}=\sqrt{gh_0}\Big{(}\kappa-\frac{1}{6}h_0^2\kappa^3\Big{)}.
    \label{KdVDR}
\end{equation}
Equation (\ref{KdVDR}) is a unidirectional approximation of (\ref{WWPDR}) that is valid in the long-wave (i.e.~$\kappa h_0\rightarrow0$) limit.  In order to obtain a weakly nonlinear approximation to (\ref{WWProb}) that is valid for a wider range of $\kappa h_0$ values, \citet{Whitham, Whithambook} proposed the following dimensional equation

\begin{equation}
    \eta_t+\sqrt{\frac{g}{h_0}}~\mathcal{K}(\kappa)*\eta+\frac{3}{2h_0}\sqrt{gh_0}~\eta\eta_x=0,
    \label{Whitham}
\end{equation}
as a model for the evolution of small-amplitude waves on shallow water.  We refer to (\ref{Whitham}) as the dimensional temporal Whitham (tWhitham) equation.  The convolution term is defined by

\begin{equation}
    \mathcal{K}(\kappa)*\eta=\tilde{\mathcal{F}}^{-1} \Big{(}\mathcal{K}(\kappa)\tilde{\mathcal{F}}(\eta)\Big{)},
\end{equation}
where $\mathcal{K}(\kappa)$ is the nondimensional Fourier multiplier given by

\begin{equation}
    \mathcal{K}(\kappa)=i~\text{sgn}(\kappa)\sqrt{\kappa h_0\tanh(\kappa h_0)},
    \label{dimK}
\end{equation}
and $\tilde{\mathcal{F}}$ and $\tilde{\mathcal{F}}^{-1}$ represent the Fourier and inverse Fourier transforms in $x$, respectively.  We write the tWhitham equation and $\mathcal{K}$ in forms different than those used by Whitham in order to simplify the work below.  The linear dispersion relation for the tWhitham equation is

\begin{equation}
    \omega_{\text{W}}=\text{sgn}(\kappa)\sqrt{g\kappa\tanh(\kappa h_0)}.
    \label{WDR}
\end{equation}
Equation (\ref{WDR}) is one root of (\ref{WWPDR}), which means that the tWhitham equation exactly models the wave speed of unidirectional linear waves for any $\kappa h_0$.

Given an initial spatial profile of the surface displacement (i.e.~given $\eta$ for all values of $x$ at $t=0$), equations (\ref{KdVDimensional}) and (\ref{Whitham}) provide predictions for the spatial profiles of $\eta$ for $t>0$.  Because of this requirement for initial spatial data, these equations are known as evolution-in-time or ``temporal'' equations.  Many laboratory and field water wave measurements are made by recording time series at fixed spatial locations (i.e.~using stationary gauges or buoys to collect time series).  Evolution-in-space or ``spatial'' equations are required to {\emph{directly}} compare with measurements of this type.  In these situations, time series at the first measurement location are used as initial conditions in spatial equations that are solved to obtain predictions for the time series at the downstream measurement locations.  Spatial equations are commonly used in nonlinear optics, e.g.~in the the study of pulse propagation in optical fibers, see for example \citet{opticswaveguides} and \citet{Optics}.

Two common methods used for deriving approximate spatial equations from equation (\ref{WWProb}) include: (i) interchanging the roles of $x$ and $t$ in the derivation of the temporal equations or (ii) starting with an approximate temporal equation and then applying the change of variables $\eta_t\sim-\sqrt{gh_0}~\eta_x$ for waves in shallow water.  Using either (i) or (ii) gives the dimensional spatial KdV (sKdV) equation

\begin{equation}
    \eta_x+\frac{1}{\sqrt{gh_0}}~\eta_t-\frac{\sqrt{gh_0}}{6g^2}~\eta_{ttt}-\frac{3}{2h_0\sqrt{gh_0}}~\eta\eta_t=0.
    \label{sKdVDim}
\end{equation}
\citet{Trillo} proposed the following dimensional spatial Whitham (sWhitham) equation

\begin{equation}
    \eta_x+\sqrt{\frac{h_0}{g}}~\mathcal{K}^{-1}(\omega)*\eta-\frac{3}{2h_0\sqrt{gh_0}}~\eta\eta_t=0,
    \label{sWhithamDim}
\end{equation}
where the convolution term is defined by

\begin{equation}
    \mathcal{K}^{-1}(\omega)*\eta=\mathcal{F}^{-1}\Big{(}\mathcal{K}^{-1}(\omega)\mathcal{F}(\eta)\Big{)},
\end{equation}
using $\mathcal{F}$ and $\mathcal{F}^{-1}$ to represent the Fourier and inverse Fourier transforms in $t$, respectively, and $\mathcal{K}^{-1}(\omega)$ as the inverse of $\mathcal{K}(\kappa)$.  Although a closed-form expression for $\mathcal{K}^{-1}(\omega)$ is not known, the inverse is guaranteed to exist since $\mathcal{K}(\kappa)$ is one-to-one and onto for $\kappa\in\mathbb{R}$.  In general, if the dispersion is monotonic and onto, then the linear initial-value problem can be written in spatial form. In other words, one can recover the spatial initial condition from time series at $x=0$ and vice versa.  Note that if weak surface tension were included, then $\mathcal{K}(\kappa)$ would not be one-to-one and a unique inverse would not exist.  

In the following, we consider both the mathematical properties of the sWhitham equation and its applications.  Section \ref{SectionProperties} includes a summary of the mathematical properties including its conserved quantities, traveling-wave solutions, and their stability.  Section \ref{SectionComparisons} contains comparisons between measurements from laboratory experiments and predictions obtained from the tKdV, sKdV, tWhitham, and sWhitham equations, and their dissipative generalizations.  Section \ref{SectionSummary} contains a summary of the results. 

\section{Properties of the spatial equations}
\label{SectionProperties}

In order to gain an understanding of the sWhitham equation, we present its properties along with the properties of the sKdV equation for comparative purposes.  We define nondimensional variables $u$, $\chi$, and $\tau$ by

\begin{equation}
    u=\frac{\eta}{h_0}, \hspace*{1cm} \chi=\frac{x}{h_0}, \hspace*{1cm} \tau=\sqrt{\frac{g}{h_0}}~t.
\label{nondimensionalization}
\end{equation}
The corresponding nondimensional sKdV equation is

\begin{equation}
    u_\chi+u_\tau-\frac{1}{6}u_{\tau\tau\tau}-\frac{3}{2}uu_\tau=0,
    \label{nondimKdV}
\end{equation}
and the nondimensional sWhitham equation is

\begin{equation}
    u_\chi+K^{-1}(w)*u-\frac{3}{2}uu_\tau=0,
    \label{sW}
\end{equation}
where $w=\sqrt{h_0/g}~\omega$ is the nondimensional circular frequency and $K$ is the nondimensional Fourier multiplier defined by

\begin{equation}
    K(k)=i~\mbox{sgn}(k)\sqrt{k\tanh(k)},
\end{equation}
where $k=h_0\kappa$ is the nondimensional wavenumber.

\subsection{Conserved quantities}
\label{SectionCQs}

The sKdV equation is known to have an infinite number of conservation laws, see for example \citet{AS}.  The first three are conservation of mass, momentum, and the Hamiltonian,

\begin{subequations}
    \begin{equation}
        \tilde{\mathcal{Q}}_1=\int_{-\infty}^{\infty}u~d\tau,
        \label{CQ1}
    \end{equation}
    \begin{equation}
        \tilde{\mathcal{Q}}_2=\int_{-\infty}^{\infty}u^2~d\tau,
    \end{equation}
    \begin{equation}
        \tilde{\mathcal{Q}}_3=\frac{1}{2}\int_{-\infty}^{\infty}\left( u^2+\frac{1}{6}u_\tau^2-\frac{1}{2}u^3\right)~d\tau,
    \end{equation}
    \label{CQ13}
\end{subequations}
respectively.  The sWhitham equation conserves mass, $\tilde{\mathcal{Q}}_1$, momentum, $\tilde{\mathcal{Q}}_2$, and its Hamiltonian

\begin{equation}
        \tilde{\mathcal{Q}}_4=\frac{1}{2}\int_{-\infty}^{\infty}\left( u\left(\frac{K^{-1}(w)}{i~w}\right)*u-\frac{1}{2}u^3\right)~d\tau.
        \label{CQ4}
\end{equation}
We emphasize that the quantities $\tilde{\mathcal{Q}}_1$ through $\tilde{\mathcal{Q}}_4$ are constant in $\chi$, the nondimensional spatial variable.  The periodic generalizations of these conserved quantities, $\tilde{\mathcal{Q}}_1$, $\tilde{\mathcal{Q}}_2$, $\tilde{\mathcal{Q}}_3$, $\tilde{\mathcal{Q}}_4$, for the tKdV, sKdV, tWhitham, and sWhitham equations are included in Appendix \ref{AppendixCQs}.

\subsection{Traveling-wave solutions}
\label{SectionTW}

The sKdV equation admits a two-parameter family of traveling-wave solutions given by

\begin{equation}
    u(\chi,\tau)=U_0+U_2~\mbox{cn}^2\left(\frac{2E_1(m)}{T}\left(\tau-\gamma\chi\right),m\right),
\end{equation}
where

\begin{subequations}
    \begin{equation}
        U_0=-\frac{16}{3T^2}E_1(m)\big{(}E_2(m)+(m-1)E_1(m)\big{)},
    \end{equation}
    \begin{equation}
        U_2=\frac{16m}{3T^2}\big{(}E_1(m)\big{)}^2,
    \end{equation}
    \begin{equation}
        \gamma=1+\frac{8}{3T^2}(m-2)\big{(}E_1(m)\big{)}^2+\frac{8}{T^2}E_1(m)E_2(m).
    \end{equation}
\end{subequations}
Here $T$ is the (temporal) period of the solution, $U_2$ is the wave height of the solution, $m\in[0,1)$ is known as the elliptic parameter of the Jacobi elliptic function $\mbox{cn}(\cdot,m)$,  and $E_1(m)$ and $E_2(m)$ are the complete elliptic integrals of the first and second kinds, respectively.  See \citet{bf} for details of elliptic functions.  The integration constant, $U_0$, was chosen so that the solutions have zero mean (i.e.~$\mathcal{Q}_1=0$) because such solutions are the most physically relevant.  This two-parameter family (the free parameters are $T$ and $m$) comprises all zero-mean, traveling-wave solutions to the sKdV equation.  A profile of the form $u(\tau-\gamma\chi)$ of the spatial equation is interpreted as the $u(-\gamma(x-{\gamma}^{-1}t))$ traveling-wave profile of the temporal evolution.  This means that the nondimensional real parameter $\gamma$ corresponds to the inverse of the nondimensional wave speed.

\begin{figure}
    \begin{center}
        \includegraphics[width=14cm]{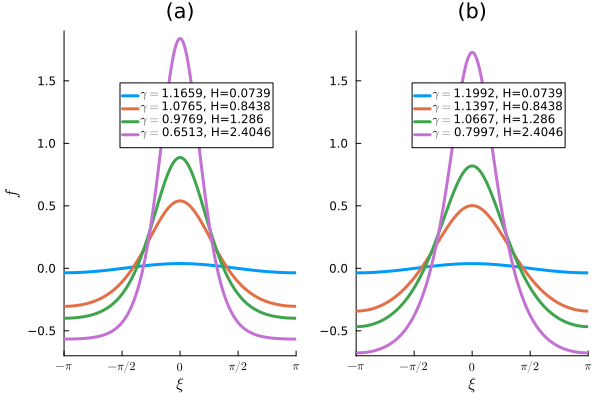}
        \caption{Plots of four $2\pi$-periodic, zero-mean, traveling-wave solutions to (a) the sKdV equation and (b) the sWhitham equation.  The $\gamma$ values and wave heights, $H$, are included in the legends.}
        \label{SolnPlots}
    \end{center}
\end{figure}

Figure \ref{SolnPlots}(a) contains plots of four $2\pi$-periodic solutions of the sKdV equation.  The wave heights, $H$, and $\gamma$ values for these solutions are included in the legend.  The sKdV equation does not admit a solution with maximal height, nor does it admit a solution with minimal $\gamma$ value.  As $m\rightarrow1$, $H$ increases without bound, and $\gamma$ decreases without bound.  Once the height of the solution becomes large enough, the $\gamma$ value becomes negative.  For solutions with period $T=2\pi$, this sign transition occurs for solutions with height $H\approx4.28$.  Since $\gamma$ is the inverse of wave speed, $\gamma$ going through zero corresponds to the wave speed going to infinity.  This nonphysical result may be due to the fact that the sWhitham equation is a generalization of the sKdV equation, which is a model of small-amplitude waves and these waves are well outside the small-amplitude regime.  

\citet{EK} proved that the tWhitham equation admits small-amplitude, periodic traveling-wave solutions and computed some of these solutions.  \citet{WhithamCusp} proved Whitham's conjecture that the tWhitham equation admits a traveling-wave solution with maximal wave height and that this solution is cusped.  \citet{NearExtremeWhitham} numerically examined the properties of solutions to the tWhitham equation close to the maximal height.

We consider periodic traveling-wave solutions of the sWhitham equation of the form

\begin{equation}
    u(\chi,\tau)=f(\tau-\gamma\chi)=f(\xi),
    \label{TWAnsatz}
\end{equation}
where $\gamma$ is a real constant and $f$ is a smooth, real-valued function of $\xi=\tau-\gamma\chi$ with nondimensional temporal period $T$.  Substituting (\ref{TWAnsatz}) into (\ref{sW}) and integrating with respect to $\xi$ once gives

\begin{equation}
    -\gamma f+\left(\frac{K^{-1}(w)}{i~w}\right)*f-\frac{3}{4}f^2=B,
    \label{TWEqn}
\end{equation}
where $B$ is the constant of integration.  This equation is invariant under the transformation

\begin{equation}
    f\rightarrow f+\nu, \hspace*{1.5cm}\gamma\rightarrow \gamma-\frac{3}{2}\nu,\hspace*{1.5cm}B\rightarrow B+\nu(\gamma-\frac{3}{4}\nu),
\end{equation}
where $\nu$ is any real constant.  Therefore, without loss of generality, we only consider traveling-wave solutions of the sWhitham equation that have zero mean.

Equation (\ref{TWEqn}) can be solved approximately by assuming $f$ has a Fourier expansion of the form

\begin{equation}
    f(\xi)=\sum_{j=-N}^{N} \hat{f}(j)\exp\left(\frac{2\pi ij\xi}{T}\right),
    \label{fForm}
\end{equation}
where $N$ is a large positive integer and the $\hat{f}$ are complex constants.  Since $f$ has zero mean, $\hat{f}(0)=0$.  For simplicity, we assume that the solutions are real and even.  Thus, $\hat{f}(-j)=\hat{f}(j)$ for $j=1,\dots,N$.  Substituting (\ref{fForm}) into (\ref{TWEqn}) gives 

\begin{equation}
    -\gamma\hat{f}(j)+\frac{TK^{-1}(\frac{2\pi w}{T})}{2\pi i~w}\hat{f}(j)-\frac{3}{4}\sum_{l=-N+j}^{N}\hat{f}(j-l)\hat{f}(l)=0,\hspace*{1cm}\text{for $j=1,2,\dots,N$.}
\end{equation}
We solved this system of nonlinear algebraic equations for the $\hat{f}$'s using Newton's method, see \citet{EK} and \citet{CVWhitham} for the details in closely related problems.

Figure \ref{SolnPlots}(b) includes plots of four $2\pi$-periodic, zero-mean, traveling-wave solutions to the sWhitham equation.  As the value of $\gamma$ decreases, the height of the solution increases.  We do not see evidence of a wave of maximum height that is analogous to the cusped wave seen in the tWhitham equation.  For the sWhitham equation, the $\gamma$ value for $2\pi$-periodic solutions becomes negative when $H\approx3.70$.  

\begin{figure}
    \begin{center}
        \includegraphics[width=12cm]{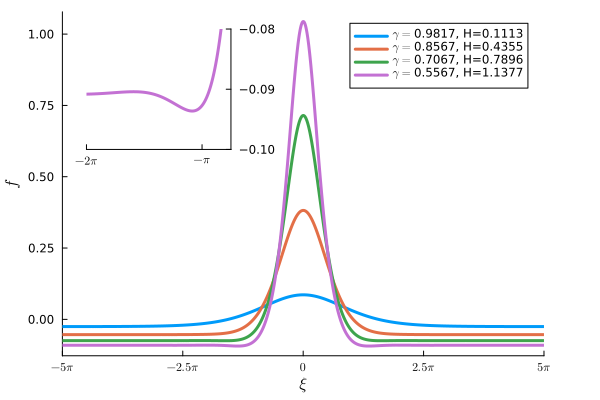}
        \caption{Plots of four traveling-wave solutions to the sWhitham equation with period $T=10\pi$.  The inset plot shows that the solutions are not monotonic on $\xi\in(-5\pi,0)$ when the solution height is sufficiently large.}
        \label{s100SolnPlot}
    \end{center}
\end{figure}

Figure \ref{s100SolnPlot} includes plots of four traveling-wave solutions to the sWhitham equation with period $T=10\pi$.  Just as in the $T=2\pi$ case, there does not appear to be a solution with maximal height, nor a solution with minimal $\gamma$ value.  However, once the height becomes large enough, the solutions no longer increase monotonically on $\xi\in(-5\pi,0)$.  This is demonstrated in the inset plot in Figure \ref{s100SolnPlot}.  This nonmonotonic behavior is not exhibited by solutions to the sKdV equation.  For clarity, we define height by $H=\max(u)-\min(u)$ regardless if the solution is monotone on $\xi\in(-\frac{T}{2},0)$ or not.

\subsection{Stability of traveling-wave solutions}
\label{SectionStability}

\citet{BenF} showed that small-amplitude periodic traveling-wave solutions to (\ref{WWProb}) are stable with respect to the modulational instability (long-wavelength perturbations) when $\frac{2\pi}{L}<1.363$ and are unstable with respect to the modulational instability when $\frac{2\pi}{L}>1.363$, where $L$ is the spatial wavelength.  \citet{BD} and \citet{DN} proved that {\emph{all}} traveling-wave solutions of the tKdV equation are stable regardless of their height or (spatial) period.  Due to the similarities between the tKdV and sKdV equations, these results also establish that all traveling-wave solutions of sKdV are stable regardless of their height or period.  \citet{HurJohnson2015} proved that traveling-wave solutions to the tWhitham equation with sufficiently small amplitude are stable with respect to the modulational instability if $2\pi/L<1.146$ and are unstable with respect to the modulational instability otherwise.  The more general work of \citet{Binswanger} also shows that the small-amplitude modulational instability cutoff occurs at $2\pi/L=1.146$ in the tWhitham equation.  \citet{Sanford2014} numerically corroborated these results, numerically showed that all periodic traveling-wave solutions of the tWhitham equation with large enough amplitude are unstable to perturbations of all wavelengths regardless of their wavelengths, and showed that the instability growth rate increases monotonically with the solution height.

Applying formula (23) of \citet{Binswanger} to the sWhitham equation establishes that all small-amplitude periodic traveling-wave solutions are stable with respect to the modulational instability regardless of their period.  This result is qualitatively similar to the tKdV and sKdV results, but is qualitatively different than the tWhitham result.

In order to numerically study the stability of traveling-wave solutions to the sWhitham equation, we employ the Fourier-Floquet-Hill method of \citet{DK}.  First, enter a coordinate frame moving with the solution via the change of variables

\begin{equation}
    \xi=\tau-\gamma\chi.
    \label{COV}
\end{equation}
This converts (\ref{sW}) to

\begin{equation}
    u_\chi-\gamma u_\xi+K^{-1}(w)*u-\frac{3}{2}uu_\xi=0,
    \label{MovingsW}
\end{equation}
and converts traveling-wave solutions of (\ref{sW}) into stationary (i.e.~$\partial_\chi u=0$) solutions of (\ref{MovingsW}).  Next, consider perturbed solutions of the form

\begin{equation}
    u_{pert}(\xi,\chi)=u(\xi)+\epsilon v(\xi,\chi)+\mathcal{O}(\epsilon^2),
    \label{PertForm}
\end{equation}
where $u$ is a periodic traveling-wave solution of the sWhitham equation, $\epsilon$ is a small real constant, and $\epsilon v$ is the leading-order part of the perturbation.  Substituting (\ref{PertForm}) into (\ref{MovingsW}) and linearizing gives

\begin{equation}
    v_\chi-\gamma v_\xi+K^{-1}(w)*v-\frac{3}{2}u^{\prime}v-\frac{3}{2}uv_\xi=0,
    \label{LinearProb}
\end{equation}
where ``prime'' means derivative with respect to $\xi$.  Without loss of generality assume

\begin{equation}
    v(\xi,\chi)=V(\xi)\mbox{e}^{\lambda\chi}+c.c.,
    \label{vForm}
\end{equation}
where $V$ is a complex-valued function, $\lambda$ is a complex constant whose real part corresponds to the growth rate of the instability, and $c.c.$ stands for complex conjugate.  Substituting (\ref{vForm}) into (\ref{LinearProb}) and rearranging gives

\begin{equation}
    \gamma V^{\prime}-K^{-1}(w)*V+\frac{3}{2}u^{\prime}V+\frac{3}{2}uV^{\prime}=\lambda V.
    \label{LinearProb2}
\end{equation}
All bounded solutions to this equation (i.e.~solutions with {\emph{any}} period) have the form $V(\xi) = e^{i \mu \xi} W(\xi)$, where $W$ is a $T$-periodic function and $\mu\in[-\frac{\pi}{T},\frac{\pi}{T}]$ is a constant known as the Floquet parameter, see \citet{DK}.  This gives

\begin{equation}
    V(\xi) = \mbox{e}^{i \mu \xi} \sum_{j=-\infty}^{\infty} {\hat W}(j)\mbox{e}^{2 \pi i j \xi /T}, 
\end{equation}
where the $\hat{W}(j)$ are complex numbers.  In our numerical computations, we use the truncation 

\begin{equation}
    V(\xi) = \mbox{e}^{i \mu \xi} \sum_{j=-N}^{N}{\hat W}(j)\mbox{e}^{2 \pi i j \xi /T},
    \label{VForm}
\end{equation}
where $N$ is a large positive integer.  If there exists a bounded solution to (\ref{LinearProb2}) with $\lambda$ that has a positive real part, then the perturbation grows exponentially in $\chi$ and the corresponding solution to the sWhitham equation is said to be unstable.  If all solutions of (\ref{LinearProb2}) have $\lambda$ values that are purely imaginary, then the corresponding solution is said to be spectrally stable.

Traveling-wave solutions to the sWhitham equation with period $T=2\pi$, height $H<0.842$, and $\gamma>1.14$ are spectrally stable.  As the wave height surpasses $H=0.842$ (and $\gamma$ falls below $1.14$), the solutions become unstable.  The spectra corresponding to solutions with heights just above the critical value are oval-like shapes centered at the origin.  The oval-like shapes correspond to perturbations with $\mu$ values near $\pm0.5$.  (A perturbation with $\mu=0.5$ has a period that is twice that of the unperturbed solution.)  As the height increases further, more $\mu$ values lead to instability and the ovals transition into figure infinities centered at the origin.  All nonzero $\mu$ values lead to instability when the complete figure infinity is formed.  For solutions with period $2\pi$, the complete figure infinity is first observed near $H=0.845$.  These sWhitham stability results are qualitatively different than the tWhitham results where the first unstable solutions have spectra with figure eights centered at the origin.  Such spectra are created by perturbations with $\mu$ values near zero and are representative of the modulational instability.

\begin{figure}
    \begin{center}
        \includegraphics[width=12cm]{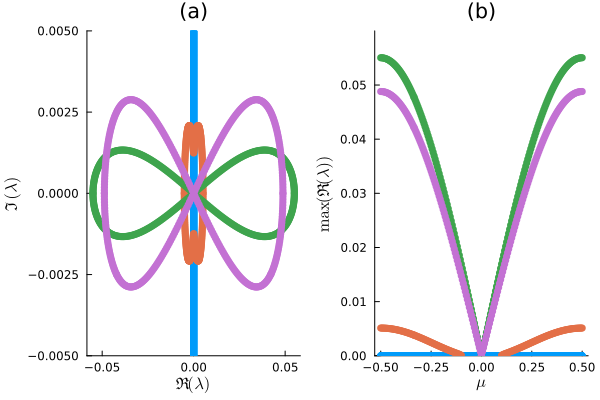}
        \caption{Plots of the stability results for the sWhitham solutions with $T=2\pi$ shown in Figure \ref{SolnPlots}(b).  The colored curves in this figure correspond to the solutions of the same colors in Figure \ref{SolnPlots}(b).  Plot (a) shows the stability spectra and (b) shows the maximum growth rate versus the Floquet parameter.}
        \label{s20StabilityPlot}
    \end{center}
\end{figure}

Figure \ref{s20StabilityPlot}(a) includes plots of the stability spectra for the sWhitham solutions shown in Figure \ref{SolnPlots}(b).  Figure \ref{s20StabilityPlot}(b) includes plots of $\max(\Re(\lambda))$, i.e.~the maximal instability growth rate, versus $\mu$ for the same solutions.  The solution with smallest height (colored blue in the plots) is spectrally stable since it has a purely imaginary spectrum.  The fact that this solution is stable is consistent with the \citet{Binswanger} asymptotic result that all traveling-wave solutions of the sWhitham equation with sufficiently small amplitudes are stable with respect to the modulational instability.  Additionally, this solution does not exhibit the ``bubble'' instabilities seen in the Euler equations, see \citet{Oliveras}, or in bidirectional generalizations of the tWhitham equation, see \citet{BernardOlga}.  The other three solutions plotted in Figure \ref{SolnPlots} are unstable.  

The solution with second smallest height (colored orange) is unstable.  Its stability spectrum is an oval-like shape centered at the origin that is transitioning into a figure infinity, see Figure \ref{s20StabilityPlot}(a).  This sWhitham solution is only unstable with respect to perturbations with approximately $|\mu|\in[0.1,0.5]$.  Since $\mu$ near zero does not lead to instability, this solution is stable with respect to the modulational instability.  The solution is most unstable with respect to perturbations with $\mu=0.5$.  This means that perturbations with period $T=4\pi$, i.e.~twice the period of the underlying solution, grow fastest.  These results are qualitatively different than those obtained for moderate-amplitude $2\pi$-periodic traveling-wave solutions of the tWhitham equation, see \citet{Sanford2014}.

The solution with the third smallest height (colored green in the plots) is unstable with respect to perturbations with any nonzero value of $\mu$.  The stability spectrum is a figure infinity centered at the origin.  The instability growth rates for this solution are larger than those for the solution with the second smallest height.  It is most unstable with respect to the perturbation with $\mu=0.5$, a perturbation with $T=4\pi$.  Finally, the solution with the largest height (colored magenta) is also unstable with respect to perturbations with any nonzero $\mu$ and its stability spectrum is a figure infinity centered at the origin.  However, the growth rates of the instabilities of this solution are smaller than the growth rates of the solution with third smallest wave height, see Figure \ref{s20StabilityPlot}(b).

\begin{figure}
    \begin{center}
        \includegraphics[width=12cm]{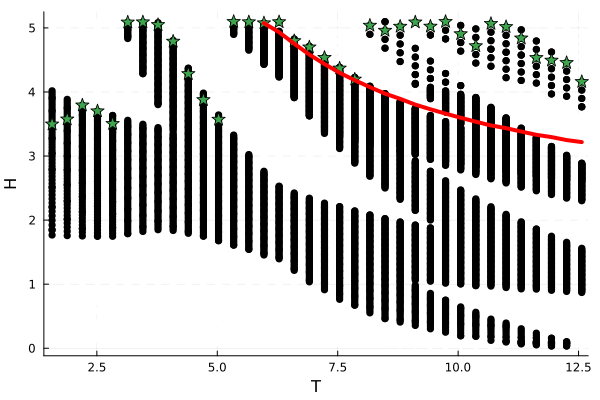}
        \caption{Plot of the regions of $(T,H)$-space for which periodic traveling-wave solutions of the sWhitham equation are unstable (black dots) with respect to the $\mu=0.5$ instability.  Solutions corresponding to $(T,H)$ in the white bands are stable with respect to $\mu=0.5$ perturbations.  The green stars represent the approximate location of the most unstable solutions for that period.   The red curve shows the approximate values of $(T,H)$ where $\gamma$ becomes negative.}
        \label{StabilitySummaryPlot}
    \end{center}
\end{figure}

As the wave height continues to increase, there are alternating bands of stability and instability.  Figure \ref{StabilitySummaryPlot} shows regions of $(T,H)$-space for which periodic traveling-wave solutions to the sWhitham equation are unstable with respect to $\mu=0.5$ perturbations (black dots).  For this plot, we only examined $\mu=0.5$ perturbations because for all $(T,H)$ pairs we examined, the $\mu=0.5$ perturbation had the largest instability growth rate.  There does not appear to be a simple relationship between $(T,H)$ and stability.  For example, all four of the $10\pi$-periodic solutions shown in Figure \ref{s100SolnPlot} are stable, but not all $10\pi$-periodic solutions are stable.  The approximate values of $(T,H)$ where $\gamma$ turns negative are shown by the red curve.  This curve does not appear to have a simple relationship with stability.  Figure \ref{StabilitySummaryPlot} shows that there are bands of instability and stability.  This is qualitatively different than what happens in the tWhitham case where all large-amplitude solutions are unstable and the growth rate increases monotonically with wave height.  It is unintuitive that some large-amplitude solutions are stable.  This unintuitive result may be attributed to the fact that the sWhitham equation is a model for small-amplitude waves and these solutions are outside of that range of validity.

Finally, we did not find any solutions to the sWhitham equation that are unstable with respect to the superharmonic instability, a perturbation that has the same period as the solution (i.e.~$\mu=0$).  Additionally, neither the momentum nor the Hamiltonian oscillate for the sWhitham equation as solution height increases.  These results are qualitatively different than the tWhitham case where traveling-wave solutions with large-enough amplitude are unstable with respect to the superharmonic instability, see \citet{SuperharmonicWhitham}.

\section{Comparisons with experiments}
\label{SectionComparisons}

The tKdV equation has been shown to compare favorably with experiments, see for example \citet{russell,zabusky,h,HS}.  \citet{Trillo} showed that both the sKdV and sWhitham equations accurately model the evolution of experimental waves of depression on shallow water.  \citet{WhithamComp} showed that the tWhitham equation more accurately predicts the evolution of experimental waves of depression than do the tKdV and Serre equations.  

In this section, we present comparisons between predictions obtained from numerical simulations of the model equations and measurements from four different series of laboratory experiments including waves of depression on shallow water (Sections \ref{SectionDepression} and \ref{SectionDissipation}), solitons on shallow water (Section \ref{SectionSoliton}), waves of depression and elevation on deep water (Section \ref{SectionUpDownDeep}), and wave packets on deep water (Section \ref{SectionDeep}).

\subsection{Numerical methods}
\label{SectionNumerics}

In order to make comparisons with measurements from laboratory experiments, the model equations need to be solved numerically.  The tKdV and tWhitham equations are solved numerically using fourth-order operator splitting in time (see \citet{yoshida}) and periodic boundary conditions in space.  The nonlinear parts of the equations are solved using fourth-order Runge-Kutta in time and a Fourier pseudospectral evaluation of the nonlinearity on a uniform grid (pointwise multiplication in space and spectral computation of the derivative).  The linear part of the equations is solved exactly in Fourier space. We use the fast Fourier transform (FFT) to move between the spatial and spectral variables.  The length of the numerical tank must be large enough that waves do not leave one end of the domain, wrap around, and impact waves on the other end of the domain.  Additionally, the spatial gridpoints must be selected to line up with the experimental gauge locations.  As a check on the results, the numerical preservation of the conserved quantities of the equations, see Appendix \ref{AppendixCQs}, was monitored.  

The sKdV and sWhitham equations are solved using the same methods except that space and time are interchanged.  The inverse $K^{-1}(w)$ is evaluated numerically using Newton's method.  As a check on the results, the preservation of the conserved quantities, see Appendix \ref{AppendixCQs}, was monitored.

\subsection{Waves of depression on shallow water}
\label{SectionDepression}

\citet{HS} conducted a series of water-wave experiments in a long, narrow tank with $h_0=10$ cm, and a wavemaker at one end.  The wavemaker was a rectangular, vertically-moving piston located on the bottom of the tank next to a rigid wall at one end of the tank.  The piston spanned the width of the tank and had a length of 61~cm, so the initial wavelength was 122~cm.  This experiment is in the shallow-water regime because the ratio $r=\frac{h_0}{L}=0.082\ll 1$ and $\tanh(2\pi r)=0.47$.  The experiments were initialized by rapidly moving the piston downward a prescribed amount.  Time series were collected by wave gauges located $61+500j$~cm for $j=0,\dots,4$ from the upstream end of the tank.  This means that the first gauge was located at the downstream edge of the wavemaker.  The tank was long enough that waves reflecting from the downstream end of the tank did not impact the time series.

We compare predictions from the tKdV, sKdV, tWhitham, and sWhitham equations.  The time series collected by the first gauge were used as initial conditions for the simulations of the spatial equations.  The initial conditions for the temporal equations require knowledge of the surface displacement for all values of $x$ at the initial time, but that information was not recorded in these experiments.  To approximate it, we used the following function as initial condition for the temporal equations

\begin{equation}
\eta(x,0)=\left\{
     \begin{array}{lr}
       0 & -7869\le x<-183,\\
       -\frac{1}{2}A_0+\frac{1}{2}A_0\mbox{sn}(0.0925434x,0.9999^2) & -183\le x\le 61,\\
       0 & 61<x\le 7747,
     \end{array}
   \right.
\label{ICs}
\end{equation}
where $A_0$ is the amplitude of the piston motion in centimeters, $x$ is measured in centimeters, and $\mbox{sn}(\cdot,m)$ is a Jacobi elliptic function with elliptic modulus $m$, see \citet{bf}.  This function represents a trough of 122~cm centered at $x=-61$~cm.  

\begin{figure}
    \begin{center}
        \includegraphics[width=12cm]{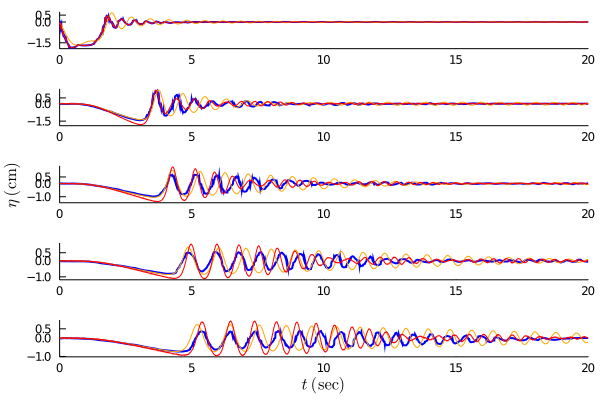}
        \caption{The experimental time series (blue curve) from the \citet{HS} experiment using $A_0=1.5$~cm with the predictions obtained from the tKdV (orange curve) and sKdV (red curve) equations.  The plots are ordered from top to bottom by increasing distance from the wavemaker.}
        \label{STKdVCompExpt3}
    \end{center}
\end{figure}

Figure \ref{STKdVCompExpt3} contains plots comparing the experimental time series with the predictions obtained from the tKdV and sKdV equations for the experiment with $A_0=1.5$~cm.  Both equations do a reasonable job predicting the experimental measurements.  However, the sKdV equation more accurately predicts the phase speed.  (This is especially visible at the downstream gauges.)  It is important to note that part of the error in the tKdV prediction is due to the fact that the initial surface is estimated.  This is a shortcoming of the temporal equations.  Both models overpredict the amplitudes at the downstream gauges and this overprediction increases as the waves travel down the tank.  This overprediction is due to the fact that both tKdV and sKdV are conservative models and the experiment contains dissipation.  See Section \ref{SectionDissipation} for a discussion of the role dissipation plays.

\begin{figure}
    \begin{center}
        \includegraphics[width=12cm]{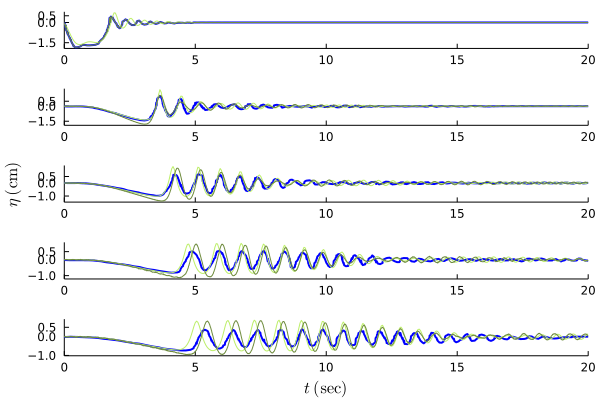}
        \caption{The experimental times series (blue curve) from the \citet{HS} experiment using $A_0=1.5$~cm with the predictions obtained from the tWhitham (light green curve) and sWhitham (dark green curve) equations.  The plots are ordered from top to bottom by increasing distance from the wavemaker.}
            \label{STWCompExpt3}
    \end{center}
\end{figure}

Figure \ref{STWCompExpt3} contains plots comparing the experimental time series with the predictions obtained from the tWhitham and sWhitham equations for the same experiment.  The differences between the tWhitham and sWhitham predictions are smaller than the differences between the tKdV and sKdV predictions.  Note that part of the error in the tWhitham prediction is due to the fact that the initial surface displacement is estimated.  Again, this highlights a shortcoming of the temporal equations.  Both the tWhitham and sWhitham equations overpredict the amplitudes of the waves at the downstream gauges due to their conservative nature; see Section \ref{SectionDissipation}.

The predictions from all four models (tKdV, sKdV, tWhitham, and sWhitham) for the \citet{HS} experiment with $A_0=0.5$~cm (plots omitted for conciseness) are more accurate than those in the $A_0=1.5$~cm case.  However, the results were qualitatively the same: the spatial equations provide more accurate predictions than do the temporal equations, the Whitham equations provide more accurate predictions than do the KdV equations, and all four equations overpredicted the wave amplitudes.  Finally, we note that linear theory is not sufficient to model the time series from either the $A_0=0.5$~cm or $A_0=1.5$~cm experiments.  Predictions obtained from linear theory (plots omitted for conciseness) are significantly worse than any of the models examined herein.

\subsection{Dissipative models of waves of depression on shallow water}
\label{SectionDissipation}

\begin{figure}[ht]
    \begin{minipage}[b]{0.45\linewidth}
    \centering
    \includegraphics[width=\textwidth]{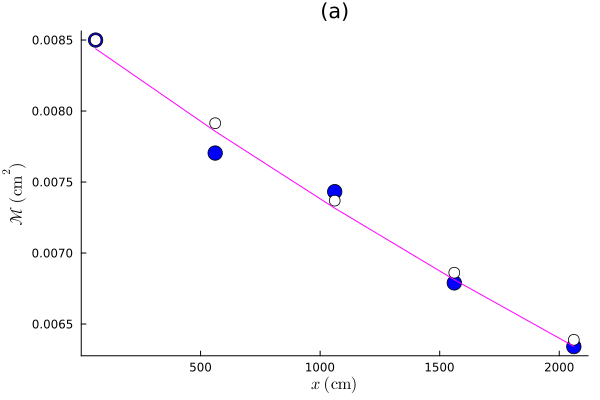}
    \end{minipage}
    \hspace{0.5cm}
    \begin{minipage}[b]{0.45\linewidth}
    \centering
    \includegraphics[width=\textwidth]{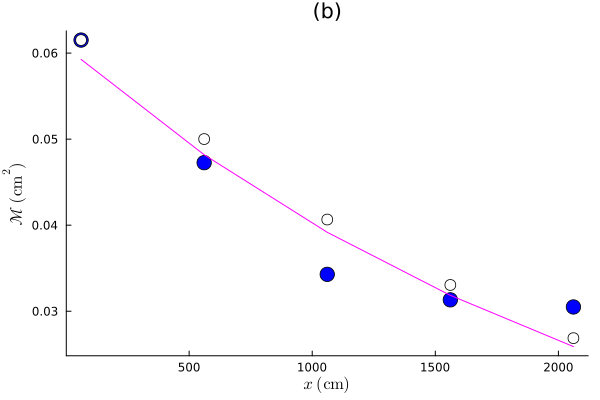}
    \end{minipage}
    \caption{Plots of $\mathcal{M}$ versus $x$ for the \citet{HS} experiments with (a) $A_0=0.5$~cm and (b) $A_0=1.5$~cm.  The blue dots represent the experimental measurements, the magenta curves represent the best-fit exponentials of the experimental data, and the black circles represent the dsWhitham predictions.}
    \label{MPlots}
\end{figure}

Figures \ref{STKdVCompExpt3} and \ref{STWCompExpt3} show that the tKdV, sKdV, tWhitham, and sWhitham equations overpredict the amplitudes of the waves measured in the experiments.  This overprediction is due to the fact that the equations are conservative, while the experiments contain dissipation.  Figure \ref{MPlots} shows that the dimensional quantity

\begin{equation}
    \mathcal{M}(x)=\frac{1}{\tau}\int_0^\tau \eta(x,t)^2~dt,
    \label{eqn:M}
\end{equation}
where $\tau$ is the dimensional length of the time series, decays nearly exponentially as the waves travel down the tank for both the $A_0=0.5$~cm and $A_0=1.5$~cm experiments.  In order to address this, we consider the dimensional dissipative sKdV equation

\begin{equation}
    \eta_x+\frac{1}{\sqrt{gh_0}}~\eta_t-\frac{\sqrt{gh_0}}{6g^2}~\eta_{ttt}-\frac{3}{2h_0\sqrt{gh_0}}~\eta\eta_t+\delta\eta=0,
    \label{SdKdV}
\end{equation}
and propose the following ad-hoc, dimensional dissipative generalization of the sWhitham equation

\begin{equation}
    \eta_x+\sqrt{\frac{h_0}{g}}~\mathcal{K}^{-1}(\omega)*\eta-\frac{3}{2h_0\sqrt{gh_0}}~\eta\eta_t+\delta\eta=0.
    \label{SdWhitham}
\end{equation}
Here $\delta$ is a nonnegative constant representing the sum total of all forms of dissipation in the experiment.  We refer to equations (\ref{SdKdV}) and (\ref{SdWhitham}) as the dissipative sKdV (dsKdV) and the dissipative sWhitham (dsWhitham) equations, respectively.  These equations predict that $\mathcal{M}$ will decay exponentially as $x$ increases (i.e.~as the waves travel down the tank).  The single free parameter, $\delta$, is determined by best fitting the measured exponential decay of $\mathcal{M}$.  The $\delta$ values for the  $A_0=0.5$~cm and $A_0=1.5$~cm experiments are $7.14*10^{-5}$ cm$^{-1}$ and $2.07*10^{-4}$ cm$^{-1}$, respectively.  This form of dissipation assumes that waves of all periods decay with the same rate.  It is not a wavenumber dependent form of dissipation.

\begin{figure}
    \begin{center}
        \includegraphics[width=12cm]{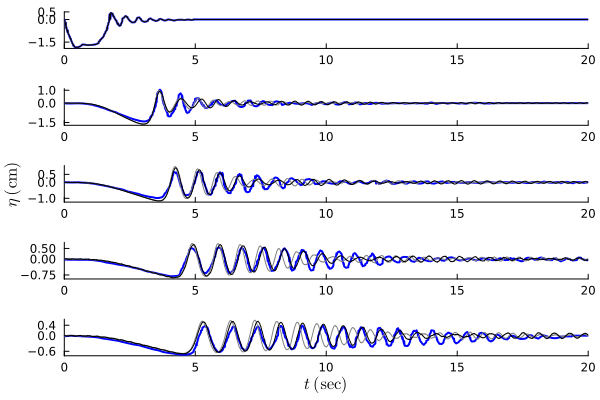}
        \caption{The experimental time series (blue curve) from the \citet{HS} experiment using $A_0=1.5$~cm and the predictions obtained from the dsKdV (gray curve) and dsWhitham (black curve) equations.}
        \label{SdWCompExpt3Plot}
    \end{center}
\end{figure}

Figure \ref{SdWCompExpt3Plot} shows comparisons between the experimental time series and predictions from the dsKdV and dsWhitham equations.  The dsKdV equation does a reasonable job, but incorrectly models the phase speeds.  The dsWhitham equation does an excellent job modeling the $A_0=1.5$~cm experimental time series.  The predictions obtained from the dsWhitham equation are much better than those obtained from the conservative models.  The results for the $A_0=0.5$~cm experiment (not shown) are similarly excellent. 

\subsection{Solitons on shallow water}
\label{SectionSoliton}

We conducted experiments on solitons in a 1,524~cm long (reflected waves did not impact the evolving soliton), 25.4~cm wide wave channel in the W.G. Pritchard Fluid Mechanics Laboratory. The tank is described in detail in \citet{vohd2017}.  Briefly,  the channel was cleaned with alcohol and filled with water to a desired depth.  A wind was blown over the water surface along the length of the channel, creating a surface current that carried surface contaminants to the other end, where they were vacuumed with a wet-vac.  The depth was then measured to be 5.30~cm.  We generated solitons using a horizontal displacement of a piston: a vertical plate that spanned the width and height of the channel.  The piston was programmed using the approach of \citet{gr1980}, to take into account the real-time displacement of the plate. This approach and the details for our wavemaker are spelled out in \citet{hhgy2004}. Herein, the wavemaker produced a repeatable KdV soliton with the desired height of 2.00~cm.  \citet{hhgy2004} also generated a KdV soliton with that height in water of depth 5.00~cm. They measured the spatial wavelength of the soliton to be about 80~cm.  We did not measure a spatial wavelength, but the ratio in the present experiments is nearby this value.  Therefore, the ratio of depth to wavelength, $r=h_0/\lambda\approx0.066$, meaning these experiments are in the shallow-water regime.  A capacitance-type wave gauge was used to measure the surface displacement.  For a fixed soliton amplitude, we conducted seven experiments with the wave gauge moved to the different $x$-locations, $x=$50, 150, 250, 350, 450, 550, 650~cm away from the piston in its rest position.  In these seven experiments, the start time of the paddle was not synced to the start time of data collection, so that the time series had to be shifted by hand to match comparisons; hence, we did not test accuracy of wave speed.

\begin{figure}
    \begin{center}
        \includegraphics[width=12cm]{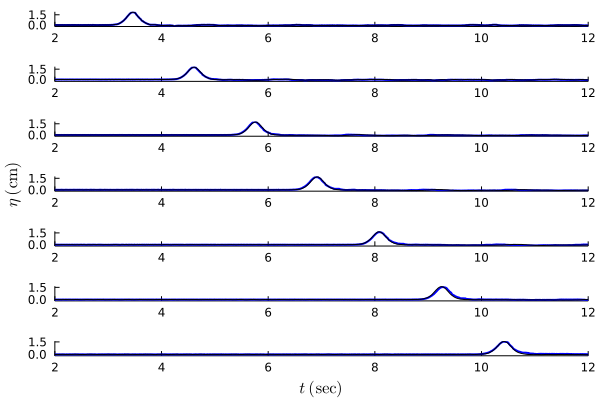}
        \caption{The experimental time series (blue curve) from the experiment described in Section \ref{SectionSoliton} along with the prediction obtained from the dsWhitham equation (black curve). The plots are ordered by increasing distance from the wavemaker.}
        \label{SdWSolitonAll}
    \end{center}
\end{figure}

\begin{figure}
    \begin{center}
    \includegraphics[width=12cm]{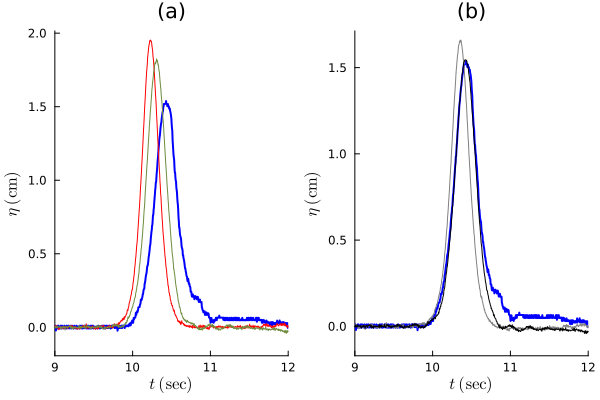}
    \caption{The experimental time series (blue curve) from the final gauge of the experiment described in Section \ref{SectionSoliton} along with the predictions obtained from (a) the sKdV (red curve) and sWhitham (dark green curve) equations and (b) the dsKdV (gray curve) and dsWhitham (black curve) equations.}
    \label{SdWSoliton7}
    \end{center}
\end{figure}

Figure \ref{SdWSolitonAll} shows a comparison of the experimental time series and the predictions obtained from the dsWhitham equation for all seven gauges.  The predictions lineup very closely with the experimental measurements.  Figure \ref{SdWSoliton7} shows comparisons of the experimental time series and the predictions from the sKdV, dsKdV, sWhitham, and dsWhitham equations for the last gauge.  Although these experiments involve waves of elevation instead of waves of depression, the results are similar to those presented in the previous two subsections.  The sWhitham predictions are more accurate than the sKdV predictions.  The dsWhitham equation more accurately models the experimental time series than did the sWhitham equation.  Other plots (omitted for conciseness) show that the dsWhitham equation is more accurate than the tKdV and sKdV equations.  Additionally, the spatial equations more accurately predict the wave evolution than do the temporal equations.

\subsection{Waves of depression and elevation on deep water}
\label{SectionUpDownDeep}

The tWhitham and sWhitham equations were proposed as models for waves on shallow water.  However, they accurately reproduce the phase speeds of all unidirectional linear waves, regardless of the (uniform) depth of water.  Because of this, it is reasonable to ask whether these equations accurately predict the evolution of waves on deep water.  

In this section, we address this question by comparing predictions from the tWhitham and sWhitham equations with time series from our experiments of waves on deep water that are similar to those discussed above in Sections \ref{SectionDepression} and \ref{SectionDissipation}.  In particular, we created an initial, localized, positive surface displacement and an initial, localized negative surface displacement. These surface displacements were created in the same tank used for the soliton experiments (see Section \ref{SectionSoliton}) but with a horizontally aligned piston that had a vertical displacement. The piston for the experiments in this section was a horizontal plate that was 25.4~cm long in the $x$-direction, 1~cm tall in the vertical direction, and spanned the width of the tank. The initial wave had a length of $\lambda=50.8$~cm, and the depth was approximately $h_0=20$~cm. The resulting ratio of fluid depth to wavelength was $r=h_0/\lambda=0.393$, with $\tanh(2\pi h_0/\lambda)=0.986$, so that the waves were effectively in the deep-water regime.

The plate was impulsively lowered or raised 0.75~cm. To obtain the motion, the operator moved a Two-Servo Joystick, which sent a signal to a Servo Travel Tuner, in which we had programmed the desired plate displacement, and then to a Hi-Tec Linear Servo, which provided the plate motion. (All parts were from www.servocity.com).  There was a vertical barrier behind the plate so that the resulting waves were forced to travel in the $x>0$ direction. 

\begin{figure}
    \begin{center}
        \includegraphics[width=12cm]{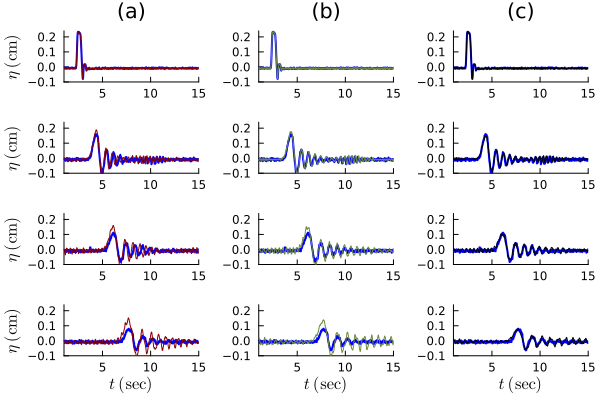}
        \caption{\label{deepD} Comparisons of predictions from the (a) sKdV equation, (b) sWhitham equation, and (c) dsWhitham equation when the initial surface displacement is positive.  The blue curves correspond to the measured surface displacement, the red, dark green, and black curves correspond to the sKdV, sWhitham, dsWhitham predictions, respectively.  The plots are ordered from top to bottom by increasing distance from the plate/wavemaker.}
    \end{center}
\end{figure}

For the positive initial displacement, the tank was filled, and the surface was cleaned as described above to a depth of 20.18~cm.  The plate was leveled parallel to the quiescent water surface and submerged 0.25~cm.  It was dropped 0.75~cm impulsively, creating a localized positive surface displacement. Four in-situ, capacitance wave gauges, located at $x=4.0$, 225.5, 461.4, 679.0~cm from the edge of the plate, provided time series.  Figure \ref{deepD} shows the results. The data from the first gauge are used as the initial conditions for the spatial equations, so the model output and measurements agree exactly at that location.  At the second gauge site, all three models predict well the oscillations that develop near the initial displacement and the wave packet that forms further downstream.  Dissipation becomes important in the surface displacement evolution by the third wave gauge. The conservative predictions agree qualitatively with the data, while the dsWhitham agrees both qualitatively and quantitatively, matching quite well even the radiation.  By the fourth gauge, the agreement between the conservative predictions and the data worsens.  The sWhitham prediction more accurately models the phase velocities than does the sKdV equation.  The dsWhitham equation continues to produce a good quantitative comparison with the data.

\begin{figure}
    \begin{center}
        \includegraphics[width=12cm]{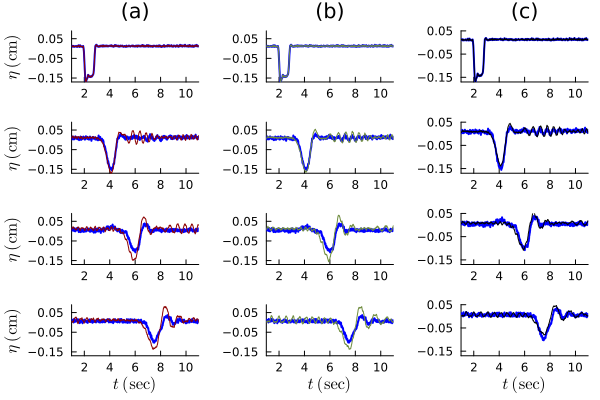}
        \caption{\label{deepU} Comparisons of predictions from the (a) the sKdV equation, (b) sWhitham equation, and (c) dsWhitham equation when the initial surface displacement is negative.   The blue curves correspond to the measured surface displacement, the red, dark green, and black curves correspond to the sKdV, sWhitham, dsWhitham predictions, respectively.  The plots are ordered from top to bottom by increasing distance from the plate/wavemaker.}
    \end{center}
\end{figure}

For the negative initial displacement, the tank was filled, and the surface was cleaned as described above to a depth of 20.22~cm.  The plate was leveled parallel to the quiescent water surface and submerged 0.75~cm.  It was lifted 0.75~cm impulsively, creating a localized negative surface displacement. The four in-situ, capacitance wave gauges, located at  $x=4.0$, 225.6, 462.5, 679.0~cm from the edge of the plate, provided time series.  Figure \ref{deepU} shows the results. The data from the first gauge is used as the initial condition, so the model output and measurements agree. Similar to the negative displacement experiments, all three models predict reasonably well the measured surface displacement at the second gauge site.  Results from the sKdV and sWhitham equations are in qualitative agreement with data at the third and fourth gauge sites, while the dsWhitham equation predicts quite well the measured time series at the third and fourth gauge sites.

The sKdV predictions are less accurate than the sWhitham predictions due to phase velocity issues
Comparisons with sKdV and sdKdV are similar, although they do not align with the experimental data as well.  

\subsection{wave packets on deep water}
\label{SectionDeep}

In order to further test the range of validity of the sWhitham and dsWhitham equations, we compare their predictions with experimental measurements of wave packets on deep water.  \citet{sh} showed that the dissipative NLS (dNLS) equation provided much more accurate predictions, both quantitatively and qualitatively, than did the NLS equation for the experiments discussed in this section.  They also showed that although plane-wave solutions to the NLS equation are unstable with respect to the modulational instability (also known as the sideband instability), uniform-amplitude solutions to the dNLS equation are stable with respect to the modulational instability.  Finally, they showed that although the dNLS equation is stable with respect to the modulational instability, the sidebands may grow a limited amount.

In the \citet{sh} experiments, the wave tank was 1,311~cm long, 25.4~cm wide, had glass sidewalls and bottom, and had a constant water depth of $h_0=20$~cm.  A plunger-type wavemaker that spanned the width of the tank, had an exponential cross-section, and oscillated vertically was located at one end of the tank.  The wavemaker created slowly modulated wavetrains with waves of wavelength $\lambda\approx15$~cm.  Since the ratio $r=h_0/\lambda\approx1.33$ and $\tanh(2\pi h_0/\lambda)\approx1.00$, this experiment is in the deep-water regime.

\begin{figure}
    \begin{center}
        \includegraphics[width=12cm]{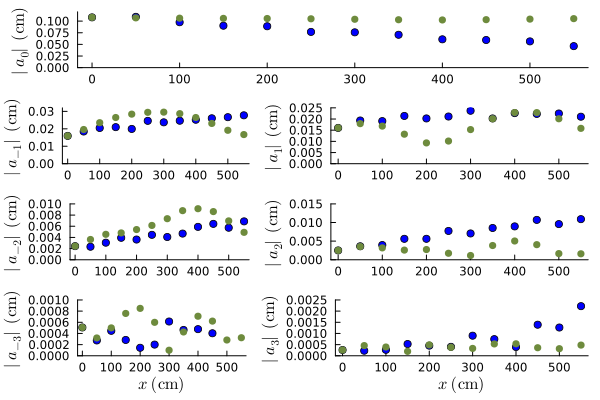}
        \caption{Plots comparing the experimental measurements (blue dots) from the deep-water experiment described in Section 6.2 of \citet{sh} with the predictions from the sWhitham equation (dark green dots).  The top plot is the carrier wave and the other six plots are the six most dominant sidebands.}
        \label{ExptASWFComp}
    \end{center}
\end{figure}

Figure \ref{ExptASWFComp} shows comparisons between the time series recorded in an experiment from Section 6.2 of \citet{sh} and the predictions from the sWhitham equation.  Due to the complexity of the time series, we compare the magnitudes of the dominant Fourier coefficients instead of the time series themselves.  The carrier wave frequency was $f_0=3.33$~Hz.  The perturbation frequency was $f_p=0.17$~Hz. So the sideband frequencies were $f_{\pm n}=f_0\pm n f_p$, for $n=1$, 2, 3. The subplots show the evolution of the carrier wave (top plot) and the six most dominant sidebands.  The plots show that the sWhitham equation does not accurately predict the evolution of these Fourier coefficients.  This discrepancy appears to be related to the modulational instability.  The experimental sidebands grow in magnitude as the waves travel down the tank (the experimental parameters were chosen so that the sidebands would grow), while the sWhitham equation predicts no such growth.  This is consistent with the fact that the sWhitham equation is not unstable with respect to the modulational instability.  An accurate model of this experimental data must allow for the sidebands to grow.

\begin{figure}
    \begin{center}
        \includegraphics[width=12cm]{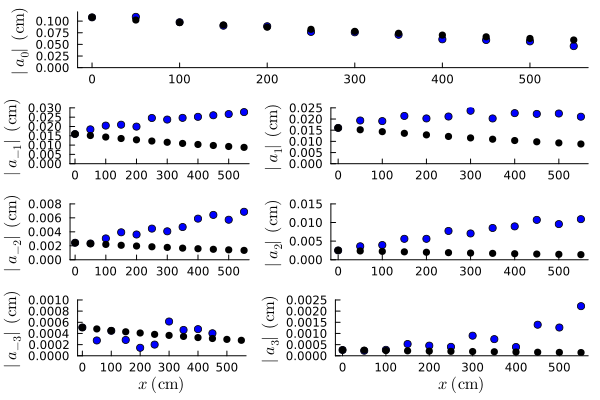}
        \caption{Plots comparing the experimental measurements (blue dots) from the deep-water experiment described in Section 6.2 of \citet{sh} with the  predictions from the dsWhitham equation (black dots).  The top plot is the carrier wave and the other six plots are the six most dominant sidebands.}
        \label{ExptASdWFComp}
    \end{center}
\end{figure}

Figure \ref{ExptASdWFComp} shows comparisons between the same experimental data and the predictions from the dsWhitham equation.  These plots show that the dsWhitham equation does not accurately model the experimental data either.  Though, note that the dsWhitham equation accurately models the evolution of the carrier wave.  Similar results are obtained when the other three deep-water experiments from \citet{FD2} are examined.  Therefore, it does not appear that the sWhitham or the dsWhitham equation can accurately model the evolution of wave packets on deep water.


\section{Summary}
\label{SectionSummary}

We examined both the mathematical properties and applications of the sWhitham equation.  The sWhitham equation can be used to model how time series of surface displacement evolve as waves travel in space. Our analysis shows that the sWhitham equation admits periodic traveling-wave solutions, but does not appear to admit a periodic traveling-wave solution with maximal height.  Small- and moderate-amplitude traveling-wave solutions to the sWhitham equation are stable with respect to the modulational instability.  Some larger-amplitude traveling-wave solutions are stable while others are unstable depending on wave period and wave height.  There does not appear to be a simple relation between wave period and height that determines stability or instability.

Our comparisons with experiments show that for waves of depression and solitons on shallow water, the sWhitham equation provides more accurate predictions for experimental time series than do the tWhitham, tKdV, and sKdV equations.  Part of the reason that the spatial predictions are more accurate than the temporal predictions is that the initial conditions need to be approximated in the temporal case.  Although the sWhitham equation was proposed as a model for shallow-water waves, we show that the sWhitham equation accurately models the evolution of initial waves of depression and elevation on deep water.  The predictions obtained from the sWhitham equation are improved by adding dissipation.  Finally, neither the sWhitham nor sdWhitham equation accurately models the evolution of wave packets on deep water.

\section{Acknowledgements}

JDC and PP thank the Dirección General de Asuntos del Personal Acad\'emico (DGAPA) at the Universidad Nacional Autónoma de M\'exico (UNAM) for a Programa de Estancias de Investigación (PREI) grant that supported JDC's sabbatical stay at UNAM.  PP acknowledges partial support by PAPIIT IG100522.  We thank Robert Geist for constructing the laboratory apparati, including the mechanical components, the motion control, and the computer interfaces, for the shallow-water soliton and all of the deep-water experiments.  We thank Vishal Vasan for writing the wavemaker program for experiments on the shallow-water soliton and the deep-water wave-packets. We thank Robert Crable for building and maintaining the capacitance wave gauges.  We thank Knox Hammack for digitizing the experimental data from \citet{HS}.  We thank Eleanor Byrnes for helpful discussions.

\appendix

\section{Conserved quantities}
\label{AppendixCQs}

The conserved quantities for the nondimensional temporal KdV equation are

\begin{subequations}
    \begin{equation}
        \mathcal{Q}_1=\int_{0}^{L}u~d\chi,
    \end{equation}
    \begin{equation}
        \mathcal{Q}_2=\int_{0}^{L}u^2~d\chi,
    \end{equation}
    \begin{equation}
        \mathcal{Q}_3=\frac{1}{2}\int_{0}^{L}\left( u^2+\frac{1}{6}u_\chi^2-\frac{1}{2}u^3\right)~d\chi,
    \end{equation}
\end{subequations}
where $L$ is the $\chi$-period of the solution.  The conserved quantities for the nondimensional temporal Whitham equation are $\mathcal{Q}_1$, $\mathcal{Q}_2$, and 

\begin{equation}
    \mathcal{Q}_4=\frac{1}{2}\int_{0}^{L}\left( u\left( \frac{K(k)}{i~k}\right)*u-\frac{1}{2}u^3\right)~d\chi.
\end{equation}
These quantities are conserved in $\tau$.

The conserved quantities for the nondimensional spatial KdV equation are

\begin{subequations}
    \begin{equation}
        \mathcal{Q}_5=\int_{0}^{T}u~d\tau,
    \end{equation}
    \begin{equation}
        \mathcal{Q}_6=\int_{0}^{T}u^2~d\tau,
    \end{equation}
    \begin{equation}
        \mathcal{Q}_7=\frac{1}{2}\int_{0}^{T}\left( u^2+\frac{1}{6}u_\tau^2-\frac{1}{2}u^3\right)~d\tau,
    \end{equation}
\end{subequations}
where $T$ is the $\tau$-period of the solution.  The conserved quantities for the nondimensional spatial Whitham equation are $\mathcal{Q}_5$, $\mathcal{Q}_6$, and 

\begin{equation}
    \mathcal{Q}_8=\frac{1}{2}\int_{0}^{T}\left( u\left( \frac{K^{-1}(w)}{i~w}\right)*u-\frac{1}{2}u^3\right)~d\tau.
\end{equation}
These quantities are conserved in $\chi$.


\begin{thebibliography}{36}
    \expandafter\ifx\csname natexlab\endcsname\relax\def\natexlab#1{#1}\fi
    \def\au#1{#1} \def\ed#1{#1} \def\yr#1{#1}\def\at#1{#1}\def\jt#1{\textit{#1}}
      \def\bt#1{#1}\def\bvol#1{\textbf{#1}} \def\vol#1{#1} \def\pg#1{#1}
      \def\publ#1{#1}\def\arxiv#1{#1}\def\org#1{#1}\def\st#1{\textit{#1}}
    
    \bibitem[Ablowitz \& Segur(1981)]{AS}
    {\sc \au{Ablowitz, M.J.} \& \au{Segur, H.}} \yr{1981} {\em Solitons and the
      Inverse Scattering Transform\/}.  \publ{SIAM, Philadelphia}.
    
    \bibitem[Agrawal(2019)]{Optics}
    {\sc \au{Agrawal, G.}} \yr{2019} {\em Nonlinear Optics\/}, {S}ixth edn.
      \publ{London: Academic Press}.
    
    \bibitem[Benjamin \& Feir(1967)]{BenF}
    {\sc \au{Benjamin, T.B.} \& \au{Feir, J.E.}} \yr{1967}  \at{The disintegration
      of wave trains on deep water: {P}art {I}. {T}heory}.  \jt{Journal of Fluid
      Mechanics}  \bvol{27},  \pg{417--430}.
    
    \bibitem[Binswanger {\em et~al.\/}(2021)Binswanger, Hoefer, Ilan \&
      Sprenger]{Binswanger}
    {\sc \au{Binswanger, A.L.}, \au{Hoefer, M.A.}, \au{Ilan, B.} \& \au{Sprenger,
      P.}} \yr{2021}  \at{Whitham modulation theory for generalized {W}hitham
      equations and a general criterion for modulational instability}.  \jt{Studies
      in Applied Mathematics}  \bvol{147},  \pg{724--751}.
    
    \bibitem[Bottman \& Deconinck(2009)]{BD}
    {\sc \au{Bottman, N.} \& \au{Deconinck, B.}} \yr{2009}  \at{{KdV} cnoidal waves
      are linearly stable}.  \jt{Discrete and Continuous Dynamical Systems A}
      \bvol{25},  \pg{1163--1180}.
    
    \bibitem[Byrd \& Friedman(1971)]{bf}
    {\sc \au{Byrd, P.F.} \& \au{Friedman, M.D.}} \yr{1971} {\em Handbook of
      Elliptic Integrals for Scientists and Engineers\/}.  \publ{Springer-Verlag,
      New York}.
    
    \bibitem[Carter(2018)]{WhithamComp}
    {\sc \au{Carter, J.D.}} \yr{2018}  \at{Bidirectional {W}hitham equations as
      models of waves on shallow water}.  \jt{Wave Motion}  \bvol{82},
      \pg{51--61}.
    
    \bibitem[Carter(2023)]{NearExtremeWhitham}
    {\sc \au{Carter, J.D.}} \yr{2023}  \at{Instability of near-extreme solutions to
      the {W}hitham equation}.  \jt{Studies in Applied Mathematics} .
    
    \bibitem[Carter {\em et~al.\/}(2023)Carter, Francius, Kharif, Kalisch \&
      Abid]{SuperharmonicWhitham}
    {\sc \au{Carter, J.D.}, \au{Francius, M.}, \au{Kharif, C.}, \au{Kalisch, H.} \&
      \au{Abid, M.}} \yr{2023}  \at{The superharmonic instability and wave breaking
      in {W}hitham equations}.  \jt{Physics of Fluids}  \bvol{35},  \pg{103609}.
    
    \bibitem[Carter {\em et~al.\/}(2018)Carter, Henderson \& Butterfield]{FD2}
    {\sc \au{Carter, J.D.}, \au{Henderson, D.M.} \& \au{Butterfield, I.}} \yr{2018}
       \at{A comparison of frequency downshift models of wave trains on deep
      water}.  \jt{Physics of Fluids}  \bvol{31},  \pg{013103}.
    
    \bibitem[Carter {\em et~al.\/}(2022)Carter, Kalisch, Kharif \& Abid]{CVWhitham}
    {\sc \au{Carter, J.D.}, \au{Kalisch, H.}, \au{Kharif, C.} \& \au{Abid, M.}}
      \yr{2022}  \at{The cubic-vortical {W}hitham equation}.  \jt{Wave Motion}
      \bvol{110},  \pg{102883}.
    
    \bibitem[Deconinck \& Kutz(2006)]{DK}
    {\sc \au{Deconinck, B.} \& \au{Kutz, J.N.}} \yr{2006}  \at{Computing spectra of
      linear operators using {H}ill's method}.  \jt{Journal of Computational
      Physics}  \bvol{219},  \pg{296--321}.
    
    \bibitem[Deconinck \& Nivala(2010)]{DN}
    {\sc \au{Deconinck, B.} \& \au{Nivala, M.}} \yr{2010}  \at{Periodic
      finite-genus solutions of the {KdV} equation are orbitally stable}.
      \jt{Physica D}  \bvol{239},  \pg{1147--1158}.
    
    \bibitem[Deconinck \& Oliveras(2011)]{Oliveras}
    {\sc \au{Deconinck, B.} \& \au{Oliveras, K.}} \yr{2011}  \at{The instability of
      periodic surface gravity waves}.  \jt{Journal of Fluid Mechanics}
      \bvol{675},  \pg{141--167}.
    
    \bibitem[Deconinck \& Trichtchenko(2015)]{BernardOlga}
    {\sc \au{Deconinck, B.} \& \au{Trichtchenko, O.}} \yr{2015}  \at{High-frequency
      instabilities of small-amplitude {H}amiltonian {PDE}s}.  \jt{Discrete and
      Continuous Dynamical Systems}  \bvol{37}~(3),  \pg{1323--1358}.
    
    \bibitem[Ehrnstr\"om \& Kalisch(2009)]{EK}
    {\sc \au{Ehrnstr\"om, M.} \& \au{Kalisch, H.}} \yr{2009}  \at{Traveling waves
      for the {W}hitham equation}.  \jt{Differential and Integral Equations}
      \bvol{22},  \pg{1193--1210}.
    
    \bibitem[Ehrnstr\"om \& Wahl\'en(2019)]{WhithamCusp}
    {\sc \au{Ehrnstr\"om, M.} \& \au{Wahl\'en, E.}} \yr{2019}  \at{On {W}hitham's
      conjecture of a highest cusped wave for a nonlocal dispersive equation}.
      \jt{Annales de l'Institut Henri Poincare. Analyse non lin\'ear}  \bvol{36},
      \pg{769--799}.
    
    \bibitem[Goring \& Raichlen(1980)]{gr1980}
    {\sc \au{Goring, D.G.} \& \au{Raichlen, F.}} \yr{1980} The generation of long
      waves in the laboratory.  \bt{In {\em Coastal Engineering Proceedings\/}}, ,
      \vol{vol.~1},  \pg{p.~46}. Sydney, Australia.
    
    \bibitem[Hammack(1973)]{h}
    {\sc \au{Hammack, J.}} \yr{1973}  \at{A note on tsunamis: their generation and
      propagation in an ocean of uniform depth}.  \jt{Journal of Fluid Mechanics}
      \bvol{60},  \pg{769--799}.
    
    \bibitem[Hammack {\em et~al.\/}(2004)Hammack, Henderson, Guyenne \&
      Yi]{hhgy2004}
    {\sc \au{Hammack, J.}, \au{Henderson, D.}, \au{Guyenne, P.} \& \au{Yi, M.}}
      \yr{2004} Solitary-wave collisions.  \bt{In {\em Proceedings of the 23rd ASME
      Offshore Mechanics and Arctic Engineering. A symposium to honor Theodore
      Yao-Tsu Wu\/}}.  \publ{Singapore: World Scientific}.
    
    \bibitem[Hammack \& Segur(1974)]{HS}
    {\sc \au{Hammack, J.L.} \& \au{Segur, H.}} \yr{1974}  \at{The {K}orteweg-de
      {V}ries equation and water waves. {P}art 2. {C}omparison with experiments}.
      \jt{Journal of Fluid Mechanics}  \bvol{65},  \pg{289--314}.
    
    \bibitem[Hur \& Johnson(2015)]{HurJohnson2015}
    {\sc \au{Hur, V.M.} \& \au{Johnson, M.A.}} \yr{2015}  \at{Modulational
      instability in the {W}hitham equation for water waves}.  \jt{Studies in
      Applied Mathematics}  \bvol{134}~(1),  \pg{120--143}.
    
    \bibitem[Johnson(2001)]{johnson}
    {\sc \au{Johnson, R.S.}} \yr{2001} {\em A Modern Introduction to the
      Mathematical Theory of Water Waves\/}.  \publ{Cambridge University Press}.
    
    \bibitem[Korteweg \& de~Vries(1895)]{kdv}
    {\sc \au{Korteweg, D.J.} \& \au{de~Vries, D.}} \yr{1895}  \at{On the change of
      form of long waves advancing in a rectangular canal, and on a new type of
      long stationary wave}.  \jt{Philosophical Magazine}  \bvol{39},
      \pg{422--443}.
    
    \bibitem[Lannes(2013)]{LannesBook}
    {\sc \au{Lannes, D.}} \yr{2013} {\em The Water Waves Problem: Mathematical
      Analysis and Asymptotics\/}.  \publ{American Mathematical Society}.
    
    \bibitem[Manakov(1974)]{opticswaveguides}
    {\sc \au{Manakov, S.V.}} \yr{1974}  \at{On the theory of two-dimensional
      stationary self-focusing of electromagnetic waves}.  \jt{Soviet Physics JETP}
       \bvol{38}.
    
    \bibitem[Miles(1981)]{miles}
    {\sc \au{Miles, J.W.}} \yr{1981}  \at{The {K}orteweg-de {V}ries equation, a
      historical essay}.  \jt{Journal of Fluid Mechanics}  \bvol{106},
      \pg{131--147}.
    
    \bibitem[Russell(1844)]{russell}
    {\sc \au{Russell, J.S.}} \yr{1844}  \at{Report on waves}.  \jt{Report of the
      fourteenth meeting of the British Association for the Advancement of Science,
      York}  \pg{pp. 311--390}.
    
    \bibitem[Sanford {\em et~al.\/}(2014)Sanford, Kodama, Carter \&
      Kalisch]{Sanford2014}
    {\sc \au{Sanford, N.}, \au{Kodama, K.}, \au{Carter, J.D.} \& \au{Kalisch, H.}}
      \yr{2014}  \at{Stability of traveling wave solutions to the {W}hitham
      equation}.  \jt{Physics Letters A}  \bvol{378},  \pg{2100--2107}.
    
    \bibitem[Segur {\em et~al.\/}(2005)Segur, Henderson, Carter, Hammack, Li,
      Pheiff \& Socha]{sh}
    {\sc \au{Segur, H.}, \au{Henderson, D.}, \au{Carter, J.D.}, \au{Hammack, J.},
      \au{Li, C.}, \au{Pheiff, D.} \& \au{Socha, K.}} \yr{2005}  \at{Stabilizing
      the {B}enjamin-{F}eir instability}.  \jt{Journal of Fluid Mechanics}
      \bvol{539},  \pg{229--271}.
    
    \bibitem[Trillo {\em et~al.\/}(2016)Trillo, Klein, Clauss \& Onorato]{Trillo}
    {\sc \au{Trillo, S.}, \au{Klein, M.}, \au{Clauss, G.F.} \& \au{Onorato, M.}}
      \yr{2016}  \at{Observation of dispersive shock waves developing from initial
      depressions in shallow water}.  \jt{Physica D}  \bvol{333},  \pg{276--284}.
    
    \bibitem[Vasan {\em et~al.\/}(2017)Vasan, Oliveras, Henderson \&
      Deconinck]{vohd2017}
    {\sc \au{Vasan, V.V.}, \au{Oliveras, K.}, \au{Henderson, D.} \& \au{Deconinck,
      B.}} \yr{2017}  \at{A method to recover water-wave profiles from pressure
      measurements}.  \jt{Wave Motion}  \bvol{75},  \pg{25--35}.
    
    \bibitem[Whitham(1967)]{Whitham}
    {\sc \au{Whitham, G.B.}} \yr{1967}  \at{Variational methods and applications to
      water waves}.  \jt{Proceedings of the Royal Society of London, A}
      \bvol{299},  \pg{6--25}.
    
    \bibitem[Whitham(1974)]{Whithambook}
    {\sc \au{Whitham, G.B}} \yr{1974} {\em Linear and Nonlinear Waves\/}.
      \publ{John Wiley \& Sons, Inc., New York}.
    
    \bibitem[Yoshida(1990)]{yoshida}
    {\sc \au{Yoshida, H.}} \yr{1990}  \at{Construction of higher order symplectic
      integrators}.  \jt{Physics Letters A}  \bvol{150},  \pg{262--268}.
    
    \bibitem[Zabusky \& Galvin(1971)]{zabusky}
    {\sc \au{Zabusky, N.J.} \& \au{Galvin, C.J.}} \yr{1971}  \at{Shallow-water
      waves, the {K}orteweg-de {V}ries equation and solitons}.  \jt{Journal of
      Fluid Mechanics}  \bvol{47},  \pg{811--824}.
    
    \end{thebibliography}
\end{document}